\documentclass[aps,prd,twocolumn,showpacs,superscriptaddress,nofootinbib,preprintnumbers]{revtex4-2}

\usepackage{comment}
\usepackage{slashed}
\usepackage{latexsym}
\usepackage{graphics}
\usepackage{bm}
\usepackage{color}
\usepackage{amsmath}
\usepackage{amssymb}
\usepackage{mathrsfs}
\usepackage{graphicx}
\usepackage{slashed}
\usepackage{siunitx}
\usepackage{braket}
\usepackage{calligra}
\usepackage{caption}
\usepackage{subcaption}

\usepackage{caption}
\usepackage{subcaption}
\definecolor{linkcolor}{rgb}{0.7752941176470588, 0.22078431372549023, 0.2262745098039215}

\usepackage{hyperref}
\hypersetup{colorlinks=true,
linkcolor=linkcolor,
citecolor=linkcolor,
urlcolor=linkcolor,
linktocpage=true
}

\def\lsim{\mathrel{\rlap{\lower4pt\hbox{\hskip1pt$\sim$}}
    \raise1pt\hbox{$<$}}}                
\def\gsim{\mathrel{\rlap{\lower4pt\hbox{\hskip1pt$\sim$}}
    \raise1pt\hbox{$>$}}}                

\newcommand{\be}{\begin{eqnarray}}
\newcommand{\ee}{\end{eqnarray}}

\newcommand{\benum}{\begin{enumerate}}
\newcommand{\eenum}{\end{enumerate}}
\newcommand{\bi}{\begin{itemize}}
\newcommand{\ei}{\end{itemize}}

\newcommand{\Eq}[1]{Eq.~(\ref{#1})}  
\newcommand{\brac}[2]{ \left( \frac{#1}{#2} \right) }

\definecolor{nicegreen}{rgb}{0.1,0.5,0.1}

\DeclareSIUnit\electronvolt{e\kern-.05em V}
\DeclareSIUnit\tonneyear{tonne-year}

\makeatletter 
    
\renewcommand\onecolumngrid{
\do@columngrid{one}{\@ne}%
\def\set@footnotewidth{\onecolumngrid}
\def\footnoterule{\kern-6pt\hrule width 1.5in\kern6pt}%
}

\renewcommand\twocolumngrid{
        \def\footnoterule{
        \dimen@\skip\footins\divide\dimen@\thr@@
        \kern-\dimen@\hrule width.5in\kern\dimen@}
        \do@columngrid{mlt}{\tw@}
}%

\makeatother 

\begin{document}

\hfill{FERMILAB-PUB-22-263-T}

\title{Constraining Feeble Neutrino Interactions
with Ultralight Dark Matter}

\author{Abhish Dev}\email{abhish@fnal.gov}
\affiliation{Theoretical Physics Department, Fermi National Accelerator Laboratory, Batavia, Illinois 60510}

\author{Gordan Krnjaic} \email{krnjaicg@fnal.gov}
\affiliation{Theoretical Physics Department, Fermi National Accelerator Laboratory, Batavia, Illinois 60510}
\affiliation{Department of Astronomy and Astrophysics, University of Chicago, Chicago, IL 60637}
\affiliation{Kavli Institute for Cosmological Physics, University of Chicago, Chicago, IL 60637}

\author{Pedro Machado}\email{pmachado@fnal.gov}
\affiliation{Theoretical Physics Department, Fermi National Accelerator Laboratory, Batavia, Illinois 60510}
\author{Harikrishnan Ramani}\email{hramani@stanford.edu}
\affiliation{Stanford Institute for Theoretical Physics,
Stanford University, Stanford, CA 94305}

\begin{abstract}
If ultralight $(\ll$ eV), bosonic dark matter couples to right handed neutrinos, active neutrino masses and mixing angles depend on the ambient dark matter density.
When the neutrino Majorana mass, induced by the dark matter background, is small compared to the Dirac mass, neutrinos are ``pseudo-Dirac" fermions that undergo oscillations between nearly degenerate active and sterile states. 

We present a complete cosmological history for such a scenario and find severe limits from a variety of terrestrial and cosmological observables. 
For scalar masses in the ``fuzzy" dark matter regime ($\sim 10^{-20}$ eV), these limits 
exclude couplings of order $10^{-30}$, corresponding to Yukawa interactions comparable to  the gravitational force between neutrinos and surpassing equivalent limits on time variation in scalar-induced electron 
and proton couplings.

\end{abstract}

\maketitle

\section{Introduction}

Ultralight $(\ll$ eV) bosonic dark matter (DM)  $\phi$ is characterized by a  macroscopic de-Broglie wavelength 
\be
\label{lambda}
\lambda_\phi = \frac{1}{m_\phi v_\phi} \approx  200 {\rm \, km} \brac{\rm neV}{m_\phi}\brac{10^{-3}}{v_\phi},
\ee
which exceeds the inter-particle separation, where $v_\phi$ is the
field veloctiy.
If $\phi$ is
misaligned from the minimum of quadratic potential, it oscillates as a classical field about this minimum according to
\be
\label{phi}
\phi(\vec r, t) = \frac{\sqrt{2\rho_{\phi}(t)}}{m_\phi} \cos [m_\phi ( t + \vec v_\phi \cdot \vec r \, ) + \varphi(\vec r\,)]~,
\ee
and the corresponding energy density redshifts like
non-relativistic matter $\rho_\phi \propto a^{-3}$, where
$a$ is the cosmic scale factor and $\varphi$ is a possible phase.
This phase may encode 
additional information about spatial variation -- e.g. different $\phi$ domains arising 
from  cosmological initial conditions\footnote{e.g. due to oscillation  starting at slightly different times in different Hubble patches when
the field becomes dynamical at $H\sim m_\phi$.} or the incoherent virialization 
in the Galaxy leading to variation on the scale of $\lambda_\phi$.

If $\phi$ couples to Standard Model (SM) particles,  
their masses, spins, and coupling constants may inherit time dependence from
 \Eq{phi}. 
In the context of charged SM particles, there are many searches for such phenomena, which typically place very strong limits
on the $\phi$-SM interaction strength (see Ref. \cite{RevModPhys.75.403} for a review). By contrast, there are relatively few bounds on
DM induced time dependence in the neutrino sector \cite{Reynoso:2016hjr,Berlin:2016woy,Krnjaic:2017zlz,Brdar:2017kbt,Davoudiasl:2018hjw,Liao:2018byh,Capozzi:2018bps,Huang:2018cwo,Farzan:2019yvo,Cline:2019seo,Dev:2020kgz,Losada:2021bxx,Huang:2021kam,Chun:2021ief} and the corresponding
limits constrain comparatively large interaction strengths primarily via flavor oscillations. 

In this paper we introduce the possibility that an
ultralight DM candidate $\phi$ induces a time dependent Majorana mass for 
right-handed neutrinos
\be
m_M = \frac{y_\phi}{2} \phi(t),
\ee
where $y_\phi$ is a coupling constant and the time dependence arises
from \Eq{phi}. 
When this mass is small compared to
the neutrino Dirac mass $m_D$, the mass eigenstates form a pair of pseudo-Dirac
fermions; one  ``active" $\nu_a$ and  one ``sterile" $\nu_s$ (per generation). These states oscillate into each other with a characteristic probability governed
by their squared mass difference $\delta m^2$ \cite{deGouvea:2004gd}
\be
\label{generic-mix}
P(\nu_a \to \nu_s) = \sin^2 (2 \theta) \sin^2\brac{\delta m^2 L}{4 E_\nu},
\ee
where $L$ is the baseline, $E_\nu$ is the energy of the
propagating neutrino, and $\theta \approx \pi/4$ is the 
mixing angle, which is near maximal in the pseudo-Dirac limit where $m_M \ll m_D$. 

The Majorana mass governing $\delta m^2$ in \Eq{generic-mix} is time dependent, so the oscillation rate becomes sensitive to the dark matter density and to its cosmic evolution. This dependence 
can impact various terrestrial and cosmological observables. 
 In this work we extract resultant bounds and impose {\it extremely} strong limits on the 
 induced Majorana mass; depending on the value of
 $m_\phi$ we find some limits on the coupling $y_\phi$ corresponding to a $\phi$ mediated Yukawa force comparable to that of gravity. 
 
 This letter is organized as follows: in section \ref{sec:theory} we present our theoretical framework, in section \ref{sec:regimes} we delineate the qualitatively different neutrino oscillation regimes that $\phi$ can induce, 
  in section \ref{sec:other} we compute the terrestrial bounds,
 in section \ref{sec:cosmo} we determine the cosmological bounds on this scenario, and in \ref{sec:conclusion} we make some concluding remarks.

\section{Ultralight dark matter and pseudo-Dirac neutrinos} \label{sec:theory}
We consider a scalar DM candidate $\phi$ with lepton number 2 and 
a cosmic abundance due to misalignment. In Weyl fermion notation, the Lagrangian 
in this scenario contains 
\be
\label{lag}
\mathcal{L}\supset y_\nu H \ell N+ \frac{ y_\phi}{2}  \phi N N  + h.c. ~,
\ee
where $y_\nu$ is the neutrino Yukawa coupling, $H$ is the SM Higgs
doublet, $\ell$ is the SM lepton doublet, and $N$ is a SM neutral fermion, i.e. a right-handed neutrino. 
As we will see next, the presence of a feeble interaction between the scalar DM and the right-handed neutrino can have dramatic effects in neutrino oscillation phenomenology.

To understand the impact of $\phi$ on neutrino oscillations, it is instructive to describe the ``1+1'' scenario, in which there is only one generation of $\ell$ and $N$.
For simplicity, assume that the active state here is an electron flavor neutrino.
In the broken electroweak phase, the first term in \Eq{lag} generates a Dirac mass of neutrinos. 
When the $\phi$ field is misaligned according to \Eq{phi}, the second
term in \Eq{lag} generates a Majorana mass for $N$, so we have 
\be
\label{masses}
m_D=\frac{ y_\nu v}{\sqrt{2}}~~,~~ m_M = \frac{y_\phi}{2} \phi(t) ~,
\ee
for the Dirac and Majorana contributions, respectively,
where $v = 246$ GeV is the Higgs vacuum expectation value.
\begin{figure*}[t]
\hspace{-1cm}
\includegraphics[width=3.4 in]{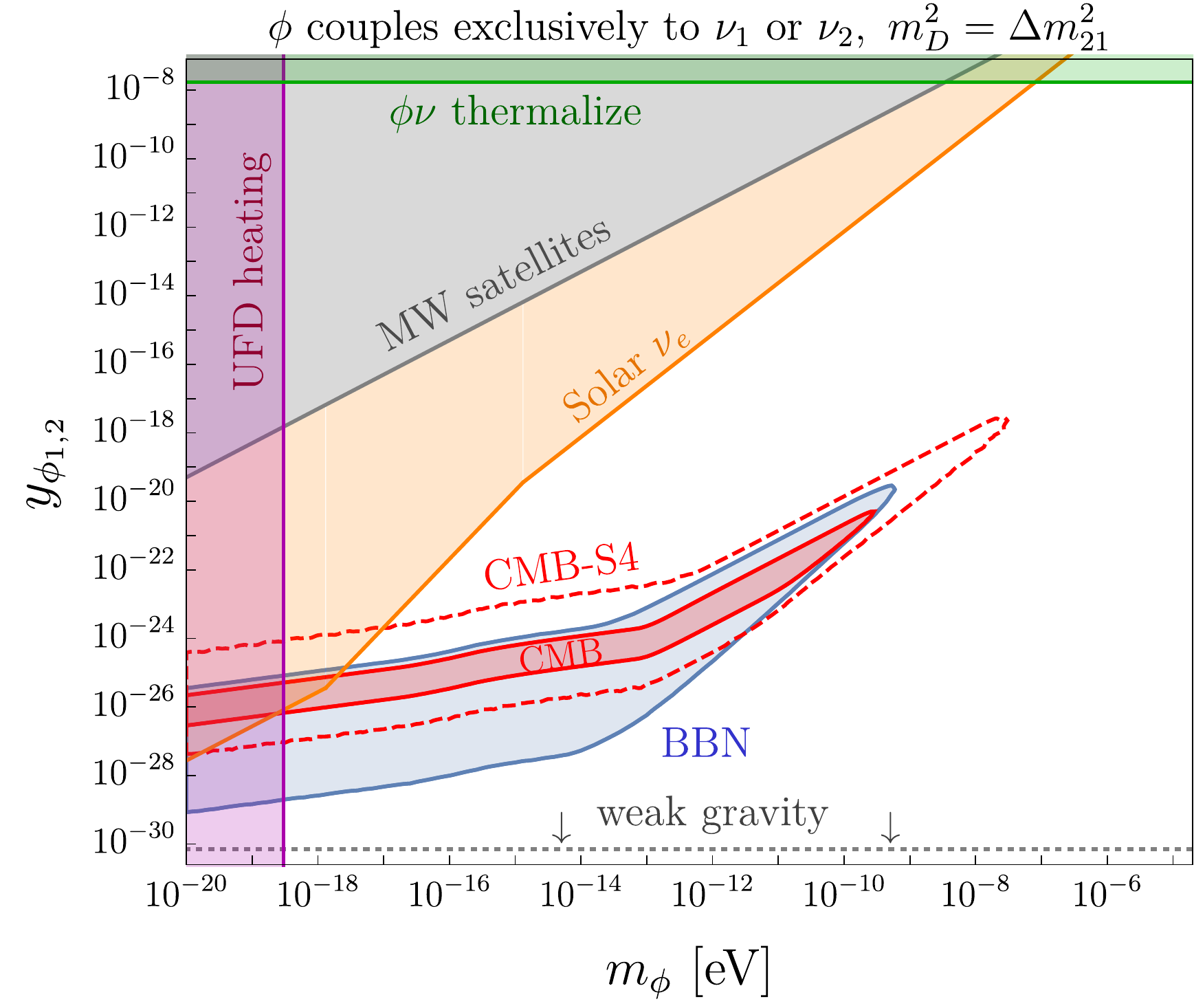}~~
\includegraphics[width=3.4 in]{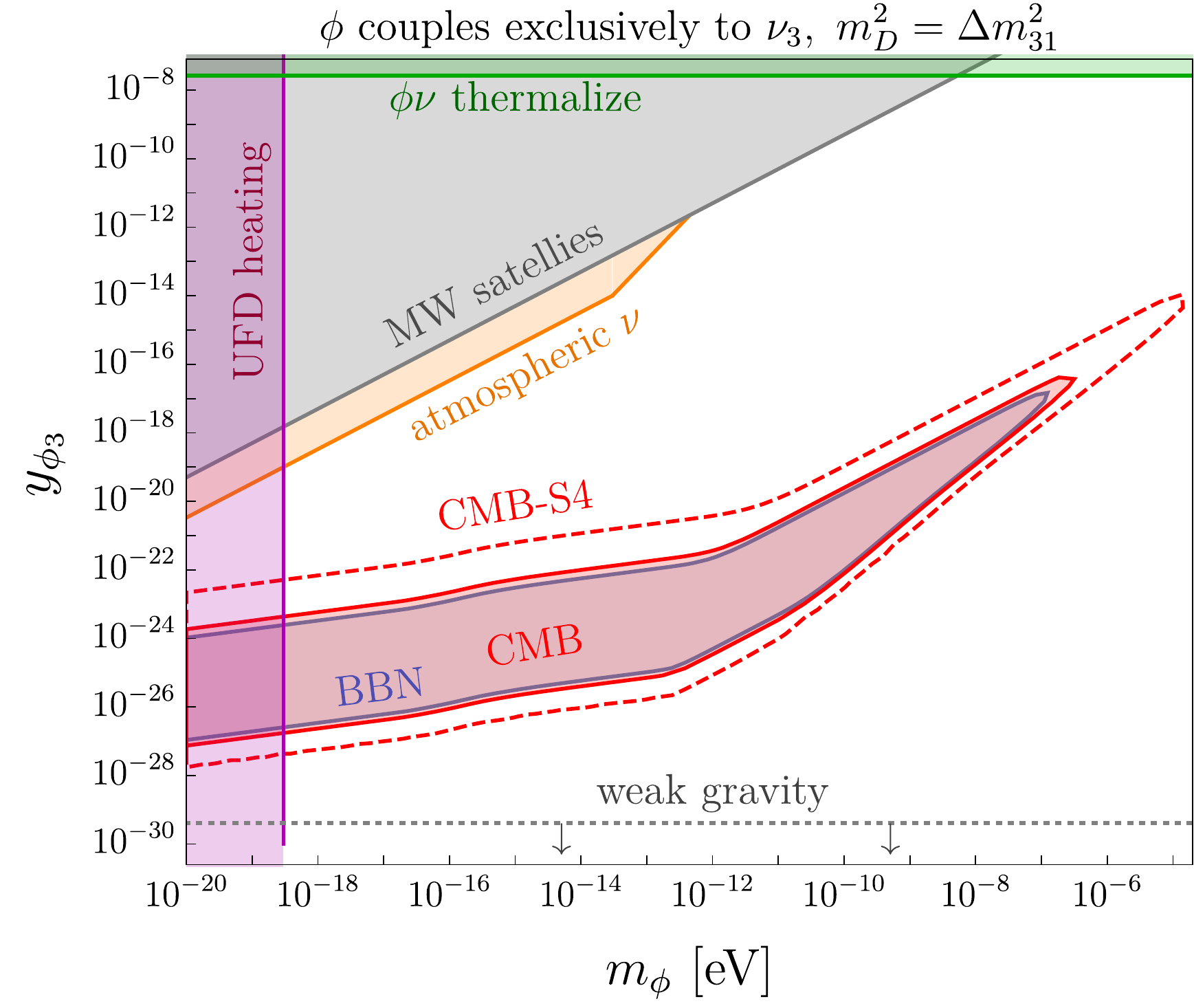}
\caption{
{\bf Left}: Parameter space for $\phi$ coupled only to
$\nu_1$ or $\nu_2$ mass eigenstates, which is predominantly
constrained $\nu_e$ oscillation bounds. Here we show bounds
from CMB and BBN from Sec. \ref{sec:cosmo}, Milky Way satellites from Sec. \ref{sec:mw},
scalar thermalization with neutrinos from Sec. \ref{sec:therm},
solar neutrino oscillations from Sec. \ref{sec:solar},
and model independent limits on light DM from
ultra faint dwarf (UFD) heating \cite{Dalal:2022rmp}. For points below the 
gray dotted line, the $\phi$ mediated force
between right handed neutrinos is weaker
than gravity, which is theoretically disfavored by the weak gravity conjecture \cite{Harlow:2022gzl}
{\bf Right:} same as the left panel, only $\phi$ now couples
only to $\nu_{3}$, so the limits are driven by $\nu_{\mu,\tau}$ oscillations for which the solar bound is
subdominant to the atmospheric bound described in Sec. \ref{sec:atm}. 
}
\label{fig:main}
\end{figure*}
When $m_M \ll m_D$, we obtain two nearly degenerate neutrino mass-squared eigenstates
\be
    m^2_{h,\ell}& = m_D^2 \pm   m_D m_M  \equiv m^2_\nu \pm \frac{1}{2} \delta m^2~,
\ee
and
we define
$\delta m^2 \equiv y_\phi m_D\sqrt{2\rho_\phi}/ m_\phi$,
where 
\be
 \delta m^2 
 \approx 2 \! \times \! 10^{-15} {\rm eV^2} \! \brac{y_\phi}{10^{-10}}\!
 \brac{10^{-15 } {\rm eV}}{m_\phi} \brac{m_D}{0.1 \, \rm eV},~~~
\ee
for the splitting between Weyl fermions as opposed to the usual $\Delta m^2_{ij}$ measured in oscillation experiments;
here we have taken the local density to be $\rho^\odot_\phi = 0.4$ GeV/cm$^3$ \cite{deSalas:2020hbh}.
The active-sterile mixing angle in this case is 
\be
\label{tan}
\tan\left(2\theta\right)=\frac{2m_D}{m_M} \gg 1 ~,
\ee
 which is nearly maximal, $\theta \approx \pi/4$ in our full parameter
 space of interest.

The diagonalization of the mass terms in \Eq{masses} is obtained by defining the flavor fields in terms of the mass eigenstates approximately as

\begin{align}
    |\nu_e\rangle=& \frac{1}{\sqrt{2}} \bigl( |\nu_h\rangle+|\nu_\ell\rangle \bigr),\\
    |\nu_s\rangle=&\frac{1}{\sqrt{2}} \bigl( |\nu_h\rangle-|\nu_\ell\rangle \bigr).
\end{align}
The time evolution of a  $\nu_e$ state is given by

\be
U(t)|\nu_e\rangle &=& 
\frac{1}{\sqrt{2}}
\biggl[
\exp\left(   - \frac{i}{2E_\nu}\int_0^t  \! dt^\prime  m_1^2(t^\prime)   \right) |\nu_1\rangle  \nonumber \\
&&
~~~ + \exp\left(   - \frac{i}{2E_\nu}\int_0^t  \! dt^\prime  m_2^2(t^\prime)   \right) |\nu_2\rangle 
\biggr],~~
\ee
which yields a $\nu_e\to\nu_e$ survival probability
\be
 P_{ee}(t)=|\langle \nu(t)|\nu_e \rangle|^2 = 
  \cos^2\left(\frac{1}{4E_\nu}\int_0^t dt^\prime \delta m^2(t^\prime)\right).~~~~~
\ee

Using Eqs.~\eqref{phi} and \eqref{masses} we obtain

\begin{equation}
\frac{1}{2} \int_0^t dt^\prime \delta m^2(t^\prime ) = \frac{ y_\phi m_D }{m_\phi} \sqrt{2 \rho_{\phi}}  \int_0^t dt^\prime \cos\left(m_\phi t^\prime+\varphi\right) ,   \nonumber
\end{equation}
where we have absorbed the $v_\phi$ dependence in 
\Eq{phi} into the definition of $\varphi$ for brevity. 
Thus, for a neutrino emitted at $t =0$ and
observed at some later time $t$, the resulting electron-neutrino disappearance probability can be written  as
\be
\label{prob}
    1\! - \!P_{ee}= 
    \sin^2 \! \left\{
        \frac{m_D} {2 E_\nu  }
    \frac{y_\phi \sqrt{2\rho_\phi}} { m^2_\phi}
    \biggl( \sin\left[m_\phi t + \varphi\right]\!-\sin\varphi \biggr)
 \!   \right\},~~~~~~
\ee
where we have treated the phase $\varphi$ as a constant over the propagation time. 

Generalization for more neutrino flavors is straightforward and can be derived following similar steps as those taken in Ref.~\cite{deGouvea:2009fp}.
Moreover, to simplify the discussion on the constraints and because the electron-neutrino admixture in $\nu_3$ is small ($|U_{e3}|\ll1$), when $\phi$ couples to $\nu_1$ or $\nu_2$ we will only consider nonstandard $\nu_e$ disappearance, while when $\phi$ couples to $\nu_3$ we will only consider nonstandard $\nu_{\mu,\tau}$ disappearance; in both regimes, we treat the active-sterile oscillation in a two-flavor (active-sterile) framework.

As written in \Eq{phi}, the phase $\varphi$ need not 
be constant over the full neutrino trajectory. Indeed,
in the Galaxy, virialization will disrupt any constant phase
value down to coherence patches of order the de-Broglie 
wavelength in \Eq{lambda}. 
Thus, the full oscillation probability
will depend crucially on the relative size
of the oscillation baseline and this coherence scale.

Finally, we note that our scalar mass is not protected by any symmetry, so it will be sensitive to irreducible one-loop corrections of order
\be
\delta m_\phi \sim \frac{y_\phi  m_D}{4\pi} \sim 10^{-18}\, {\rm eV} \brac{y_\phi}{10^{-15}} 
\brac{m_D}{10 \, \rm meV},~
\ee
from the interactions in \Eq{lag}. Thus, for small $y_\phi$ in the pseudo-Dirac limit, this contribution does not destabilize the 
ultralight scalar mass, assuming no $\phi$ couplings to heavier states.\footnote{The operator $k H^\dagger H |\phi|^2$ is also allowed by all symmetries and can induce a large correction to $m_\phi$ if the coefficient is not suppressed. Exponential $k \ll 1$ suppression can be achieved in UV models where $H$ and $\phi$ are localized on different branes in a higher dimensional spacetime.}

\section{Neutrino Oscillation Regimes}
\label{sec:regimes}
In what follows, we will consider three distinct regimes for neutrino oscillations in the presence of the ultralight scalar fields. 
These regimes arrive from the relation between the neutrino oscillation length and the modulation frequency of $\phi$ or the coherence length that defines the overall phase $\varphi$.
Instead of performing a detailed fit of experimental data, we will recast existing constraints on pseudo-Dirac neutrinos from Ref.~\cite{Cirelli:2004cz} on our parameters of interest, $y_\phi$ and $m_\phi$. 
As neutrinos are ultra-relativistic, we identify $t=L$ in \Eq{prob}.

\subsection{Constant $\phi$: $ m_\phi L \lesssim 1 $}
\label{sec:short}
In the low frequency $m_\phi L \lesssim 1$ regime, the neutrino encounters a
constant phase  $\varphi$ domain over the course of its
propagation.

Expanding \Eq{prob} around $m_\phi L\to0$ yields an oscillation
probability

\be
\label{short}
    1-P_{ee} \approx
     \sin^2 \left(
       \frac{L}{4E_\nu}
       \frac{2y_\phi  m_{D}}{m_\phi }\sqrt{2\rho_\phi}\cos\varphi
      \right).
\ee
We can interpret this oscillation probability as follows. 
Since the period of the field $\phi$ is too long compared to the neutrino time-of-flight, the pseudo-Dirac mass splitting induced by the field is constant for each neutrino.
Nevertheless, as an experiment collects data, the mass splitting will evolve as the field $\phi$ displays time modulation.
In practice, several neutrino experiments have a high enough rate of events to observe time modulation of oscillation probabilities with periods as short as 10 minutes, which would correspond to $m_\phi\sim 10^{-18}$~eV~\cite{Berlin:2016woy, Krnjaic:2017zlz, Brdar:2017kbt, Dev:2020kgz}.

Since any small pseudo-Dirac mass splitting leads to maximal mixing, time modulation of neutrino oscillation probabilities due to $\phi$ modulation would lead to large, observable  effects on oscillation data.

Both constant and time dependent pseudo-Dirac mass splittings would be ruled out by neutrino data if observed, and can be used to set limits on 
the coupling strength $y_\phi$ for a given $m_\phi$. 
Since sterile neutrino
oscillation constraints are typically reported
as bounds on $\delta m^2$, we can
define an effective mass-squared $\delta m^2_{\rm eff}$ by equating the arguments of 
\Eq{generic-mix} and \Eq{short} to obtain
\be
\label{meff}
\delta m_{\rm eff}^2  \equiv
\frac{2 y_\phi m_{D}
 }{m_\phi }\sqrt{2\rho_\phi}~,
\ee
assuming $\cos\varphi\sim1$.
Recasting pseudo-Dirac
neutrino limits on $\delta m^2$ in \Eq{meff}
allows to constrain 
\be
    \label{ylim-short}
    y_{\phi} < \frac{m_\phi}{2 m_D} 
    \frac{\delta m^2_{\rm lim} }{\sqrt{2\rho_{\rm \phi}}},
    \label{eq:limsum1}
\ee
where we have identified  $\delta m_{\rm eff}^2$ with the constrained value $\delta m_{\rm lim}^2$. 

Note that, depending on context, 
$\rho_{\phi}$ can either be the cosmological
DM density at a given cosmic era  
or the present day local density.

\subsection{Modulating $\phi$:  $(m_\phi v_\phi L <  1 \ll m_\phi L)$ }
\label{sec:mid}

When the $\phi$ modulating frequency is high, $ m_\phi L \gg 1$, the accumulated phase due to propagation is sufficient to induce many modulation cycles on $\phi$ over the neutrino trajectory. 
However, as long as $m_\phi v_\phi L \lesssim 1$, the neutrino time-of-flight is shorter than separation time of  $\phi$ wave packets. 
A neutrino propagating
in this regime will encounter the same value of $\varphi$ across its trajectory, that is, the modulation of $\phi$ throughtout the neutrino trajectory is coherent. 
Without loss of generality, we can set the initial condition $\varphi = 0$.
The effective oscillation probability in this regime is given by a time-average of \Eq{prob} over the duration of propagation
\be
    \langle 1-P_{ee}\rangle 
    \approx
    \sin^2 \left(\frac{ y_\phi m_D } {2E_\nu  m^2_\phi}
    \sqrt{2\rho_\phi}
    \right),
\ee
where we have assumed that $\rho_\phi$ does not
change appreciably across the baseline. 
In this intermediate regime we repeat the argument leading 
up to \Eq{ylim-short} and constrain
    \be
    \label{ylim-med}
   y_\phi^{\rm lim} = \frac{\delta m^2_{\rm lim} m_\phi^2 L}{2 m_D\sqrt{2\rho_{\phi}} }~. 
    \ee

\subsection{Random walk: $1 \ll m_\phi v_\phi L$}
\label{sec:long}

Finally, in the $m_\phi v_{\phi} L \gg 1$ 
regime, the neutrino time-of-flight is longer than
the wave packet separation of $\phi$, so the neutrino traverses
a random sample of $\phi$ field patches, each
with a different phase $\varphi$.
Along this trajectory, there are approximately $ m_\phi v_{\phi} L$ patches whose contributions add incoherently, so the effective
phase can be approximated by $\varphi_{\rm eff}\sim \sqrt{m_\phi v_\phi L}$, assuming random distribution of phases $\varphi$ and the phase averaged probability
can be written
\begin{align}
    \langle 1-P_{ee}\rangle \approx
        \sin^2 \left(\frac{ y_\phi m_D 
    \sqrt{ 2 \rho_\phi  v_{\phi } L }} {2E_\nu  m^{3/2}_\phi}\right).
\end{align}
The corresponding limit on
the coupling reads 
\be
    y_\phi^{\rm lim}=
    \frac{\delta m^2_{\rm lim} }{m_D} 
    \sqrt{ \frac{ m_\phi^3 L} {2\rho_\phi v_\phi}  }.
    \label{ylim-long}
\ee

\section{Terrestrial Observables}
\label{sec:other}
We now consider various
terrestrial bounds on pseudo-Dirac
neutrinos in the context of our 
scenario. Depending on the values of 
 $y_\phi$ and $m_\phi$, a particular constraint can apply in any of the 
 three regimes outlined in Sec. \ref{sec:regimes}, so the relationship between $y_\phi$ and $m_\phi$ will differ in each case

\subsection{Solar Neutrinos}
\label{sec:solar}
For electron neutrinos, 
the pseudo-Dirac splitting can be constrained
by measurements of the solar neutrino
flux. 
In the standard three neutrino oscillations paradigm, $^8$B neutrinos undergo an adiabatic evolution due to large matter effects in the Sun~\cite{Parke:1986jy}.
This leads to a survival probability $P(\nu_e\to\nu_e)\simeq c_{13}^4s_{12}^2+s_{13}^4\simeq0.3$, where $s_{ij}$ and $c_{ij}$ are the sines and cosines of mixing angle $\theta_{ij}$.
Low energy solar neutrinos, on the other hand, are not affected by matter effects, and thus 
$P(\nu_e\to\nu_e)\simeq c_{13}^4(c_{12}^4+s_{12}^4)+s_{13}^4\simeq0.55$.
These probabilities describe well experimental data~\cite{Super-Kamiokande:2005wtt, Super-Kamiokande:2008ecj, SNO:2009uok, Super-Kamiokande:2010tar, Bellini:2011rx, SNO:2011hxd, KamLAND:2013rgu, Borexino:2013zhu, Super-Kamiokande:2016yck}.
This can be used to extract an order of magnitude bound
on the splitting in our scenario by demanding 
that this prediction is not affected by an order one amount. Here we use $\rho_\phi^\odot$ and $v_\phi \approx 10^{-3}$ with $L = 1.5 \times 10^8$~km, which requires 
\be\label{solar}
\delta m^2_{\rm lim} <  10^{-12}\, \rm eV^2~~,
\ee
and can be translated into a bound on our model parameters using the relations in Sec. \ref{sec:regimes}, where the appropriate regime is
determined by $m_\phi$. 
Since solar neutrinos are essentially almost pure $\nu_2$ or incoherent $\nu_e$, and $\nu_e$ has but a small admixture $\nu_3$ mass eigenstate,
the corresponding solar limit on the $y_\phi$ applies only to the right-handed partners $N_{1,2}$.
Applying the solar limit from \Eq{solar} 
to the three regimes from Sec. \ref{sec:short}, and assuming that the Dirac mass of $\nu_1$ satisfies
$m_D^2 = \Delta m_{21}^2 = 7.4 \times 10^{-5}$ eV$^2$ \cite{Zyla:2020zbs}, we find the following constraints.

For $m_\phi \lesssim 10^{-18}$ eV, 
solar neutrinos are in the constant $\phi$ regime,
so from \Eq{ylim-short}, we find a limit
\be
    y^{\rm lim}_{\phi} \approx 3 \times 10^{-26} \brac{m_\phi}{10^{-18} \textrm{eV}},~~{\rm for}~m_{\phi}<10^{-18}~{\rm eV}.~~~~~
\ee
Note that if $m_\phi\lesssim 10^{-24}$~eV, the period of $\phi$ is larger than 20 years, and the observation of pseudo-Dirac mass splittings become dependent of the initial condition $\varphi$.
For $10^{-18} \, \text{eV} \lesssim m_\phi \lesssim 10^{-15}$ eV, we are in the modulating $\phi$ regime where \Eq{ylim-med} yields a limit of order
\be
    y^{\rm lim}_{\phi}
    \approx 3 \times 10^{-26} \brac{m_\phi}{10^{-18} \textrm{eV}}^{1/2}\!,~~{\rm for}~m_{\phi}<10^{-15}~{\rm eV}.~~~~~~
\ee
Finally, for $m_\phi \gtrsim 10^{-15}$ eV, solar neutrinos will traverse a random sample of phases $\varphi$, corresponding to the random walk regime, so the bound from
\Eq{ylim-long} applies to give
\be
        y^{\rm lim}_{\phi}
    \approx  2 \times 10^{-20}\brac{m_\phi}{10^{-15} \, \rm eV }^{3/2}\!,~~{\rm for}~m_{\phi}>10^{-15}~{\rm eV}.~~~
\ee
These results are plotted in the left panel
of in Fig. \ref{fig:main}, which shows
constraints on $\phi$ coupled only
to $\nu_1$ or $\nu_2$, corresponding
to $\nu_e$ oscillations measurements. 

\subsection{Atmospheric neutrinos}
\label{sec:atm}
Measurements of the atmospheric neutrinos can
place limits on the $\phi$ coupling to $\nu_3$ since muon neutrinos have a large admixture of the $\nu_3$ eigenstate.
If $\nu_3$ is split in a pseudo-Dirac pair, a substantial deficit of atmospheric $\nu_\mu$ flux would be observed, contradicting experimental data~\cite{Super-Kamiokande:1998kpq, Super-Kamiokande:2010orq, Super-Kamiokande:2015qek, Super-Kamiokande:2017yvm}. 
 The characteristic atmospheric baseline is the Earth's radius  $L \approx 6000$~km, 
 and the Super-Kamiokande  constraint on constant
 pseudo-Dirac mass splittings 
 is~\cite{Cirelli:2004cz}
 \be
 \label{delta-atm}
\delta m^2_{\rm lim} <  10^{-4} \, \rm eV^2, 
 \ee
 which translates into a bound
 on the $\phi$ coupling to $\nu_3$. For ultralight $\phi$ masses, atmospheric oscillations
 are in the constant $\phi$ regime of Sec.~\ref{sec:short}, so translating the constraint from \Eq{delta-atm}
 with Dirac mass satisfying $m^2_D = 
 \Delta m_{32}^2 = 2.4 \times 10^{-3}$ eV$^2$ \cite{Zyla:2020zbs} yields
 \be
    y_{\phi}^{\rm lim} \approx  10^{-14} \brac{m_\phi}{3 \times 10^{-14} \textrm{eV}}~,~~
\ee
which is valid for $m_\phi \lesssim 3 \times 10^{-14}\, \rm eV$.
For larger $\phi$ masses in the modulating $\phi$ regime of Sec. \ref{sec:mid}, we impose the limit 
\be
    y_3^{\rm lim} \approx 10^{-14} \brac{m_\phi}{3 \times 10^{-14} \, \rm eV}^{2}~.
\ee
These bounds are presented in the orange shaded region of 
Fig. \ref{fig:main} (right panel). 
Note that for  $m_\phi \gtrsim 10^{-10}$ eV, atmospheric oscillations are in the long baseline regime of Sec. \ref{sec:long}, but the bound in this mass
range is subdominant to other constraints
in Fig. \ref{fig:main} and is not shown. 
In principle atmospheric neutrinos also bound the $\phi$ coupling to $\nu_{1,2}$, but solar constraints are stronger.
 
\section{Cosmology}
\label{sec:cosmo}
\subsection{Scalar Evolution}

Throughout our analysis, we assume that the $\phi$ potential
can be written as 
\be
V(\phi) = m_\phi^2 |\phi|^2 + \frac{\lambda_\phi}{4} |\phi|^4 + {\cal O}(|\phi|^6)~
\ee
  where, in principle, the size of the quartic is unconstrained by symmetry arguments and can take on any value. However, there is an irreducible contribution to the quartic interaction generated through a Coleman-Weinberg
interaction with the neutrinos 
\be
\lambda_{\phi}^{\rm min} \approx \frac{y_\phi^4}{16\pi^2},
\ee
which is always present in the absence of fine tuning.
In an expanding Friedmann-Robertson-Walker universe, $\phi$ satisfies
the equation of motion
\be
\ddot \phi + 3 H\dot \phi + V^\prime=0~.
\ee
where the prime denotes a derivative with respect to $\phi$.

If $\phi$ is initially displaced from its minimum, it is frozen
by Hubble friction until $H\dot\phi \sim V^\prime$, so if the mass
term dominates the potential, $V^\prime \sim m_\phi^2 \phi$, the field
becomes dynamical when $m_\phi \sim H$ and oscillate about $\phi = 0$
while redshifting like non-relativistic matter $\rho_\phi \propto a^{-3}$.

In this scenario, the initial misalignment amplitude $\phi_i$ during inflation sets the DM abundance. Since $m_\phi/H_i \ll 1$, where $H_i$ is the Hubble scale during inflation, $\phi$ generates isocurvature perturbations, which are constrained by CMB measurements \cite{Planck:2018vyg}. However, as long as $\phi_i/H_i \gg 1$, isocurvature perturbations can be parametrically suppressed, so for a given $H_i$, 
a suitable choice of $\phi_i$ can account for the 
DM abundance while being safe from this constraint. 
Furthermore, since $m_\phi/H_i \ll 1$, $\phi$
evolution is predominantly classical during inflation, so the initial amplitude $\phi_i$ remains
 a free parameter throughout our analysis and can be chosen
 to yield the observed DM density \cite{Tenkanen:2019aij}.

\subsection{Milky Way Satellites}
\label{sec:mw}
In order for $\phi$ to account
for the full DM abundance, it must redshift
like non-relativistic matter $(\rho_\phi \propto a^{-3})$ in the early
universe, starting at least at matter-radiation
equality at a critical redshift $z_\star \sim  10^6$, corresponding to a temperature $T_{\star} \sim$ keV \cite{Das:2020nwc}. Since the $\phi NN$ interaction in
\Eq{lag} yields an irreducible
quartic scalar self-interaction term,
we need to ensure that the 
$\rho_\phi$ is not
dominated by the quartic contribution
at $T_{\rm eq}$; otherwise it 
would redshift like radiation
$\rho_\phi \propto a^{-4}$ (or faster if even higher 
polynomial terms dominate instead) \cite{Turner:1983he}. Avoiding
this fate requires 
\be
 m_\phi^2 |\phi_\star|^2 > \frac{y_\phi^4}{16\pi^2} |\phi_\star|^4 ~~,~~  
\ee
where $\phi_\star \equiv \phi(T_\star)$. Using the scaling in \Eq{phi}, we find
\be
\label{alwaysNR}
\hspace{1cm}
y_\phi \lesssim \left[ \frac{8\pi^2  m_\phi^4 }{\Omega_{\rm dm} \rho_{\rm c}} 
\brac{T_{0}}{T_{\star}}^{3/2}
\right]^{1/4} \!\!\! \approx  5 \times 10^{-9} \brac{m_\phi}{ \rm neV},~~~~~~
\ee
where $\rho_c$ is the present day critical density. The inequality 
in \Eq{alwaysNR} defines the gray shaded 
regions in Fig. \ref{fig:main} where this effect would
erase the Milky Way satellites already observed. However, note that this bound is model-dependent 
as it can be evaded if $\phi$ is only a small fraction
of the total dark matter abundance, in which case it need not redshift
like nonrelativistic matter at early (or even later) times. 

\subsection{Avoiding Thermalization}
\label{sec:therm}

The Yukawa interaction in \Eq{lag} enables $\phi \nu \to \phi \nu$
scattering which can bring $\phi$ particles in the misaligned
condensate into equilibrium with neutrinos if the rate ever
exceeds Hubble expansion. Since active neutrinos don't couple
directly to $\phi$, the cross section for this process
requires two Dirac mass insertions and 
scales as $\sigma \sim y_\phi^4 m_D^2/T^4$.
Furthermore,
since both $\phi$ and the neutrinos are ultralight, the scattering rate scales as $\Gamma \sim n\sigma \propto T^{-1}$, up until $T\sim m_D$ corresponding
to the maximum rate relative to Hubble. Demanding that less than
$3.8 \% $  of the $\phi$ population in the condensate 
is upscattered and becomes relativistic \cite{Poulin:2016nat} at this  temperature implies 
\be
y_\phi \lesssim \brac{ 0.038\sqrt{g_\star} m_D }{m_{\rm Pl}}^{1/4} \approx    10^{-8} \brac{10 \, \rm meV}{m_D}^{1/4},~~~~~
\ee
where $g_\star = 3.36$. This bound is shown in Fig. \ref{fig:main} as the green shaded region. 

\subsection{CMB/BBN}

 In this section, we investigate the effects of the scalar field in the early universe, specifically active to sterile oscillations, which increase
 the effective number of neutrino species,
  $\Delta N_{\rm eff}$.
 
 \subsubsection{Cosmological Field Density}
 If the relic density was set by the misalignment mechanism, then the DM density grows as $T^3$ and remains as non-relativistic DM until the temperature $T_H$ when $m_\phi= 3 H(T_H)$, where $H$ is the Hubble parameter. Above this temperature, the field is constant due to Hubble friction and only 
 contributes to the vacuum energy, so we have 
\begin{equation}
\label{rho-phi-full}
    \rho_\phi(T) = \rho_\phi(T_0) \brac{g_{\star,S}(T_0)}{g_{\star,S}(T)}\left(\frac{\min\left(T,T_H\right)}{T_0}\right)^3,
\end{equation}
where $T_0 = 2.72$ K is the present day CMB temperature and $\rho_\phi(T_0) = \Omega_{\rm dm} \rho_{c}$ is the cosmological DM density, which is related to the local overdensity via
$\rho_\phi(T_0) \approx 3 \times 10^{-6} \rho_\phi^\odot$.
In what follows, we insert \Eq{rho-phi-full} into $m_M(T) = y_\phi \phi(T)/2$ using \Eq{phi} to model the 
Majorana mass as function of cosmic temperature.

\subsubsection{Cosmological Sterile Neutrino Production}

To compute the  early universe sterile neutrino yield, it is convenient to define $r_\beta$ as the ratio
of active/sterile momentum moments 
\be r_\beta \equiv \frac{ \langle p^\beta \rangle_s }{\langle p^\beta \rangle_a}~,
\ee
where angular brackets $\langle \cdots \rangle_{s,a}$ denote a thermal
average over the sterile and
active distributions, respectively. Generalizing the formalism of 
  Ref. \cite{Dodelson:1993je}, $r_\beta$ 
  satisfies the Boltzmann equation
\begin{equation}
\label{eq:boltz}
    \frac{dr_\beta}{dT}=
    -\frac{1}{2 H T \langle p^\beta \rangle_a}\int \frac{d^3p}{(2\pi)^3}
    \frac{p^\beta \Gamma  \sin^2(2\theta_M)  }{e^{p/T}+1}   ,
\end{equation}
where
   $ \Gamma=\frac{7\pi}{24}  G_F^2 p T^4,$ and the mixing angle is 
\be
\label{thetaM}
    \sin^2 (2\theta_M) = \frac{\sin^2(2\theta_0)}{\left[\cos (2\theta_0) -2 p     V_{\rm eff}/\Delta m^2 \right]^2+ \sin^2 (2\theta_0)},~~~~~~
\ee
where the effective matter potential for each flavor $a = e,\mu,\tau$ can be written as
\be
    V^a_{\rm eff}= \pm C_1 \eta G_F T^3 - \frac{C^a_2}{\alpha} G_F^2  T^4 p, 
\ee
where $\eta = (n_L-n_{\bar L}) /n_\gamma = 6\times 10^{-10}$ is the lepton asymmetry, 
$C_1 = 0.95$, $C^e_2 \approx 0.61$, $C_2^{\mu,\tau} \approx 0.17 $, and the $\pm$
refer to neutrinos and antineutrinos ~\cite{Dolgov:2003sg}. Here the vacuum
mixing angle $\theta_0$ in \Eq{thetaM} is $\phi$ dependent
\be
\label{theta-cosmo}
\theta_0 = \tan^{-1}\brac{y_\phi \sqrt{2\rho_\phi}}{m_D m_\phi} ,
\ee
where we have used Eqs.~(\ref{masses}) and (\ref{tan}).  Note that the first two moments of
the active neutrino distribution yield
the number and energy densities ($\langle p^0\rangle_a = n_a, \langle p^1\rangle_a = \rho_a$)

In the following subsections, we derive detailed $\Delta N_{\rm eff}$ limits from BBN and CMB based on $\nu_a \to \nu_s$ oscillations around $T\approx \textrm{MeV}$; later oscillations do not affect light element yields or the Hubble rate. 
The oscillation probability is maximized when the argument of \Eq{generic-mix} is order one, implying
\be
      \frac{\delta m^2 L}{T} \sim \frac{m_D m_M}{G_F^2 T^6} 
      \sim
      \brac{y_\phi}{10^{-29}}
      \brac{m_D}{10 \, \rm meV }\brac{10^{-12}\, \rm{eV}}{m_\phi},~~~~~~
\ee
where we have used $\delta m^2 = m_D m_M$ and $m_M \sim y_\phi \phi$
from \Eq{masses}, $\phi \sim \sqrt{\rho_\phi}/m_\phi \propto T^{3/2}$ from \Eq{phi},
 and  approximated $L \sim (G_F^2T^5)^{-1}$ 
as the neutrino mean-free-path, setting $T =$ MeV throughout. Thus, 
the blue-shifted DM density
at BBN greatly enhances the 
neutrino Majorana mass 
and yields on order-one oscillation 
probability for {\it extremely} feeble
couplings $y_\phi \sim {\cal O}(10^{-29})$.
Note that in our numerical study below, we 
use the full temperature dependence from
\Eq{rho-phi-full} which also accounts
for the Hubble damped regime when $T > T_H$ and is relevant for the smallest values of $m_\phi$ we consider.

However, from \Eq{rho-phi-full}, for sufficiently large values of $y_\phi$ and $\rho_\phi$,
 $m_M > m_D$
so neutrinos are no longer pseudo-Dirac fermions at high temperatures. In this regime, $\nu_a \to \nu_s$ oscillations
 are sharply suppressed as $\theta_0 \to \pi/2$ in
 \Eq{theta-cosmo}, so there is a ceiling to the couplings
  that can be probed in the early universe; this effect yields 
  concave regions for the BBN/CMB regions in Fig. \ref{fig:main}.

\subsubsection{Extracting the CMB $\Delta N_{\rm eff}$ limit}
For temperatures before active
neutrino decoupling, sterile neutrinos produced via $\nu_a \to \nu_s$ oscillations  contribute to $\Delta
N_{\rm eff}$, which can be constrained using Cosmic Microwave Background anisotropy data.
Oscillations that take place after neutrino decoupling interchange active and sterile states, but do not contribute
to $\Delta N_{\rm eff}$.
In terms of $r$ in \Eq{eq:boltz}, sterile production via $a$ flavor oscillations predicts  
\be
\label{eq:neff-cmb}
\Delta N^{\rm CMB}_{\rm eff} = r_1\left(T_{\rm dec}^{\nu_a}\right)
,
\ee
where $T_{\rm dec}^{\nu_e} \approx 3.2 $ MeV and $T_{\rm dec}^{\nu_\mu,\nu_{\tau}} \approx 5.34$ MeV are the
temperatures of $\nu_a \bar \nu_a \to e^+e^-$ chemical decoupling \cite{Dolgov:2003sg}.
Assuming the $\Lambda$CDM cosmological model, the Planck collaboration
constraints $\Delta N_{\rm eff} \lesssim 0.28$ \cite{Planck:2018vyg} and we show 
this constraint in Fig. \ref{fig:main} as the blue shaded region alongside
projections from future measurements
with CMB-S4 \cite{CMB-S4:2016ple}.

\subsubsection{BBN $\Delta N_{\rm eff}$ Limit}
A nonzero $\Delta N_{\rm eff}$ from sterile production also yields a larger initial
neutron/proton fraction at the onset of BBN, which increases the primordial helium
fraction. As in \Eq{eq:neff-cmb}, for $\phi$ coupled to $\nu_{\mu,\tau}$, the effect on
BBN  arises purely from the expansion rate via
\be
\label{eq:neff-bbn1}
\Delta N^{\rm BBN}_{\rm eff} = r_1(T_{\rm dec}^{\nu_{\mu,\tau}}),
\ee
where $r_1$ is the solution to \Eq{eq:boltz} with $\beta = 1$ evaluated at decoupling, assuming no initial population of steriles. 
The blue contour of Fig. \ref{fig:main} (right panel) shows parameter space where $\Delta N^{\rm BBN}_{\rm eff} > 0.5$ \cite{Blinov:2019gcj,Gariazzo_2022} for  
$\phi$ coupled to the $\nu_3$ mass eigenstate, implying oscillations from $\nu_\mu$ and $\nu_\tau$ flavor states. 

However, for $\nu_e \to \nu_s$ oscillations, there are two distinct effects that impact the $n/p$ ratio: oscillations before
$\nu_e$ chemical decoupling at $T_{\rm dec}^{\nu_e} \approx 3.2$ MeV change the expansion rate as above, and oscillations after
decoupling {\it deplete} the $\nu_e$ density. Both effects can be captured
with a shift in the effective Fermi
constant via 
\be
G_F^2 \to \frac{1}{2}G_F^2 [2  
+ r_2(T_{\rm dec}^{\nu_e})
-r_2(T_{\rm nuc})
 ], 
\ee
and a simultaneous shift in $g_\star$ via  
\be
g_{\star} \to 
g_{\star, \rm SM} + \frac{7}{4} r_1(T_{\rm dec}^{\nu_e}) ~~,
\ee
where $g_{\star, \rm SM} = 10.75$ during BBN and
 $T_{\rm nuc} \approx 0.8$ MeV is the temperature
 at which nucleon inter-conversion freezes out in the SM. Note that $\langle p^2 \rangle_a \propto T^5$, which sets the 
weak scattering rate $\Gamma \sim G_F^2 T^5$, so $r_2 = \langle p^2 \rangle_s/\langle p^2 \rangle_s$ yields the fractional departure from this rate.

 We can economically capture both effects
with an equivalent $\Delta N^{\rm BBN}_{\rm eff}$
\cite{Barbieri:1989ti} to obtain
\be
\label{eq:neff-bbn2}
\Delta N^{\rm BBN}_{\rm eff} \approx 
r_1(T_{\rm dec}^{\nu_e}) + 
\frac{4}{7} g_{\star, \rm SM } \left[ r_2(T_{\rm nuc}) - r_2(T_{\rm dec}^{\nu_e}) \right].~~~~ 
\ee
 In Fig. \ref{fig:main}
 (left panel) the blue shaded region shows the BBN exclusion
 for which $\Delta N^{\rm BBN}_{\rm eff} > 0.5$. 
 
\section{conclusions}
\label{sec:conclusion}
In this letter we have presented the first cosmologically viable 
model in which neutrino masses acquire time dependence through
their coupling to ultralight dark matter. In our scenario,
the DM interaction sets the right-handed neutrino Majorana mass
and neutrinos are pseudo-Dirac fermions with small mass splittings 
between active and sterile states. 
Since in the pseudo-Dirac regime the mixing angle between active and sterile is maximal,

we extract limits  on {\it ultra feeble} Yukawa couplings between
DM and right-handed neutrinos, constraining values 
of order $y_\phi \sim 10^{-30}$ for $m_\phi \sim 10^{-19}$ eV in
the ``fuzzy" DM regime \cite{Hu:2000ke}; for such small couplings the $\phi$ mediated Yukawa force between right-handed neutrinos is comparable to that of gravity.

Throughout our analysis, we have emphasized bounds from solar and atmospheric neutrino oscillations, large scale structure, and the CMB/BBN eras. However,
additional limits on this scenario may also be extracted by studying cosmic ray propagation \cite{deGouvea:2009fp} or diffuse supernova background 
neutrinos \cite{deGouvea:2022dtw,Beacom:2010kk}, which we leave for future work.

\bigskip
\begin{acknowledgments}

This work is supported by the Fermi Research Alliance, LLC under Contract No. DE-AC02-07CH11359 with the U.S. Department of Energy, Office of Science, Office of High Energy Physics. Harikrishnan
Ramani acknowledges the support from the Simons Investigator Award 824870, DOE Grant DE-SC0012012,
NSF Grant PHY2014215, DOE HEP QuantISED award
no. 100495, and the Gordon and Betty Moore Foundation Grant GBMF7946. 
This work was performed in part at the Aspen Center for Physics, which is supported by NSF Grant No.~PHY-1607611. This project has received support from the European Union’s Horizon 2020 research and innovation programme under the Marie Skłodowska-Curie grant agreement No 860881-HIDDeN.
\end{acknowledgments}

\bibliography{biblio}

\end{document}